\documentclass[a4paper]{article}

\usepackage{INTERSPEECH2020, amsmath,graphicx,multirow,cite,url,booktabs,subfigure}
\usepackage{algorithm}
\usepackage{comment}
\usepackage[noend]{algpseudocode}
\usepackage{adjustbox}
\usepackage[flushleft]{threeparttable}
\usepackage{ifthen}

\title{Masked Proxy Loss For Text-Independent Speaker Verification}
\name{Jiachen Lian$^*$, Aiswarya Vinod Kumar$^*$, Hira Dhamyal$^*$, Bhiksha Raj$^+$, Rita Singh$^+$}
\address{
  $^*$Electrical and Computer Engineering, $^+$Language Technologies Institute\\
  Carnegie Mellon University\\
  Pittsburgh, PA, USA - 15213}
\email{\{jlian2, avinodku\}@andrew.cmu.edu, \{hyd, bhiksha, rsingh\}@cs.cmu.edu}

\begin{document}

\maketitle
\begin{abstract}
Open-set speaker recognition can be regarded as a metric learning problem, which is to maximize inter-class variance and minimize intra-class variance. Supervised metric learning can be categorized into pair-based learning and proxy-based learning \cite{Proxyanchor}. Most of the existing metric learning objectives belong to the former division, the performance of which is either highly dependent on sample mining strategy or restricted by insufficient label information in the mini-batch. Proxy-based losses mitigate both shortcomings, however, fine-grained connections among entities are either not or indirectly leveraged. This paper proposes a Masked Proxy (MP) loss which directly incorporates both proxy-based relationship and pair-based relationship. We further propose Multinomial Masked Proxy (MMP) loss to leverage the hardness of speaker pairs. These methods have been applied to evaluate on VoxCeleb test set and reach state-of-the-art Equal Error Rate(EER).
\end{abstract}

\section{Introduction}
\label{sec:intro}
Text-independent speaker verification \cite{text-independent} is the task of verifying speakers' claimed identities from recordings of their voice with no expectation that the words spoken in two recordings are the same.
The most successful approaches are to use the deep neural network to derive discriminative speaker embeddings. These approaches typically take one of two forms. 

In the first approach, the network is trained as a multi-class classification network that {\em classifies} a large number of speakers using softmax crossentropy loss and its enhanced variants \cite{metric, L-softmax,L-GM-1,L-GM-2,center-loss,sphereface}. The expectation is that this behavior will generalize to data outside the training set. 

The second approach, which relates directly to the topic of this paper, is based on {\em metric learning}. In metric learning, instead of explicitly modelling the classes, the training losses are computed directly from {\em comparison of distances} between instances. A number of  metric learning objectives have been proposed for this purpose, such as the Contrastive   \cite{voxceleb2,metric}, Triplet \cite{triplet, metric}, GE2E \cite{GE2E,metric}, Prototypical \cite{Prototypical, metric}, and Angular Prototypical \cite{metric} losses. 

 While this approach has proven to be more effective than the classification-based approach \cite{metric}, it faces considerable logistic challenges. Firstly, in order to maximally ensure that the distance between embeddings of same-class instances is consistently smaller than that between instances of different classes, ideally the distances (or similarities) between {\em every} pair of same-class instances must be compared to every pair of either of those instances with every recording from the {\em other} classes in the training set. 
 For a training set of size $N$ (recordings), raw computation of such comparisons scales by $O(N^3)$, an infeasible computation for any reasonably-sized training corpus.

Secondly,  model updates are usually performed over {\em minibatches} of data rather than the entire training set, and most metric learning approaches restrict themselves to comparisons of classes that are present within the minibatches \cite{GE2E,Prototypical,metric}.  While the computation here is less daunting -- scaling only by $O(NB^2)$ for minibatches of size $B$ for a single pass through the data -- a different suboptimality results.  The size of minibatches is typically much smaller than the total number of classes in the training data. Even with exhaustive comparison of all triplets within each minibatch,  updates made to the model will not have considered the complete possible set of comparisons across classes.


Proxy NCA \cite{Proxy-NCA}, attempts to address these deficiencies by introducing learnable {\em proxy} embeddings, one per class in the training set, and replacing instance-instance comparisons with instance-proxy comparisons, thereby reducing the time complexity to visit all comparisons to $O(NP)$, where $P$ is number of proxies. Since each training instance  is compared to {\em all} proxies, including those for classes not currently in the minibatch, Proxy-NCA leads not only to faster convergence, but potentially also better models. However, by concentrating on entity-to-proxy distances (by ``entity'', we mean an actual data instance, as opposed to the proxies which are purely synthetic), it loses the more fine-grained entity-to-entity relationships. Proxy Anchor \cite{Proxyanchor} attempts to address this by assigning dynamic weights to the entity-to-proxy comparisons,  however, the real entity-to-entity relationships remain unexploited, since the actual comparisons are still between entities and proxies.


This paper proposes a Masked Proxy(MP) loss which takes both entity-to-entity distance and entity-to-proxy distance into consideration. Within each minibatch we utilize entity-to-entity comparisons for all classes represented  in the minibatch, computed through the comparison of a randomly reserved query sample to batch centroids for the within-minibatch classes, and utilize entity-proxy comparisons only for classes not represented in the minibatch (i.e. by masking out the proxies for the within-minibatch classes). To leverage multi-similarity \cite{MSLOSS} and the hardness of positive pairs \cite{Proxyanchor}, we also propose a Multinomial Masked Proxy (MMP) loss. To the best of our knowledge, this is the first work to apply proxy-based loss in speaker recognition. Using our proposed approach, we achieve state-of-art Equal Error Rates(EER) on the VoxCeleb test set training on the VoxCeleb2 dev set.
\section{Related Works}

\subsection{Proxy-Based Loss}

In contrast to pair-based metric learning which computes losses entirely from authentic data instances, proxy-based losses utilize synthetic, learnable proxies 
to compose triplets. 


\subsubsection{Proxy NCA}
Proxy NCA \cite{Proxy-NCA} assigns a proxy to each data point according to its class label. 
The objective is to make each embedding closer to its proxy than other proxies, as shown below:
\begin{equation}\label{nca}
    L_{NCA}=-\frac{1}{N}\sum_{i=1}^{N}\mathrm{\log}(\frac{e^{-d(x_i,p_i)}}{\sum_{j=1,j\neq i}^{N}e^{-d(x_i,p_j)}})
\end{equation}
where $d$ denotes Euclidean distance.
\vspace{-0.2cm}

\subsubsection{Proxy Anchor}
In contrast to Proxy NCA, Proxy Anchor \cite{Proxyanchor} regards the proxy as an {\em anchor} that instances are drawn to. Both positive and negative pairs contribute to the loss objective by their hardness, which is illustrated in Equation 6 in \cite{Proxyanchor}.

\begin{equation}\label{anchor}
    \begin{split}
        L_{Anchor}&=\frac{1}{\left | P_+ \right |}\sum_{\substack{p\in P_+ }}\mathrm{\log}(1+\sum_{x\in X_p^+}e^{-\alpha(s(x,p)-\delta)})\\&+\frac{1}{\left | P \right |}\sum_{\substack{p\in P }}\mathrm{\log}(1+\sum_{x\in X_p^-}e^{\alpha(s(x,p)+\delta)})
    \end{split}
\end{equation}
Here $P$ represents the proxy set, $P_{+}$ represents the set of proxies for all classes represented in the minibatch, $X_p^+$ denotes the set of all instances (from the minibatch) belonging to the same class as proxy $p$, and $X_p^-$ denotes set of instances {\em not} from the same class as $p$. $\alpha$ is a scaling factor, $\delta$ is a margin and $s$ refers to cosine similarity. 

Note that neither Proxy NCA nor Proxy Anchor invoke actual entity-to-entity distances, thus not leveraging the fine-grained distance/similarity relations within a minibatch.

\section{Proposed Masked Proxy Methods}
\label{sec:loss}
\begin{figure*}[!ht]
\centering
    \begin{minipage}{.24\textwidth}
        \begin{subfigure}
        \centering
        \includegraphics[width=\textwidth]{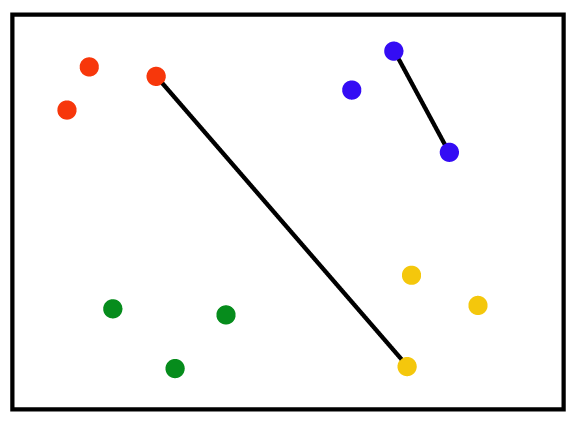}
        \caption{Contrastive}
        \label{fig:1}
        \end{subfigure}
        \begin{subfigure}
        \centering
        \includegraphics[width=\textwidth]{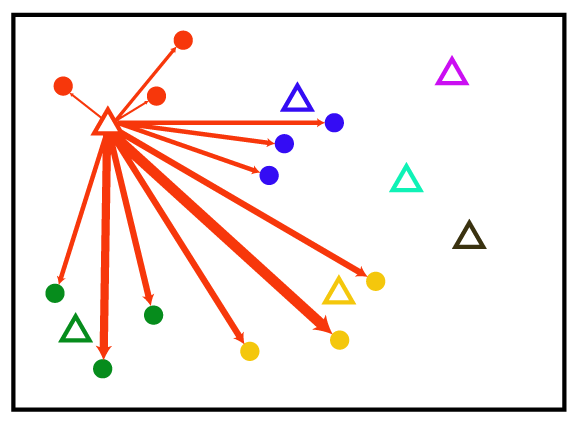}
        \caption{ProxyAnchor}
        \label{fig:4}
        \end{subfigure}%
    \end{minipage}
    \hfil
    \begin{minipage}{.24\textwidth}
        \begin{subfigure}
        \centering
        \includegraphics[width=\textwidth]{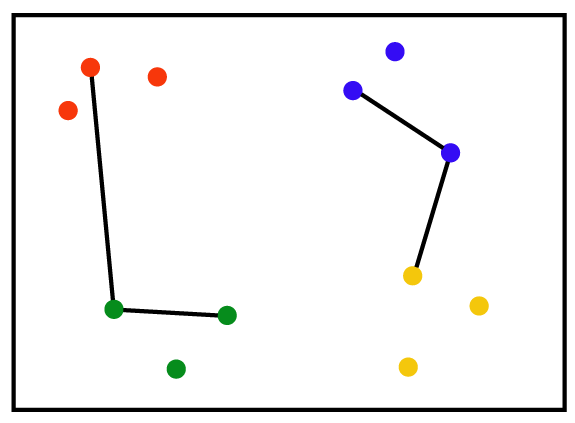}
        \caption{Triplet}
        \label{fig:2}
        \end{subfigure}
        \begin{subfigure}
        \centering
        \includegraphics[width=\textwidth]{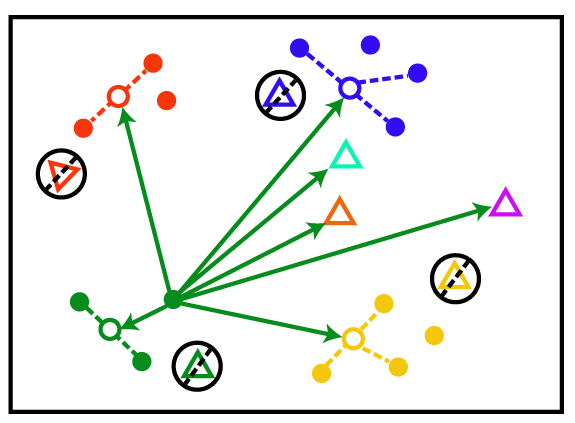}
          \caption{MP l1 loss(Eq. \ref{eq:minibatchmaskproxy})}
          \label{mp_l1}
        \end{subfigure}%
    \end{minipage}
    \hfil
    \begin{minipage}{.24\textwidth}
        \begin{subfigure}
        \centering
        \includegraphics[width=\textwidth]{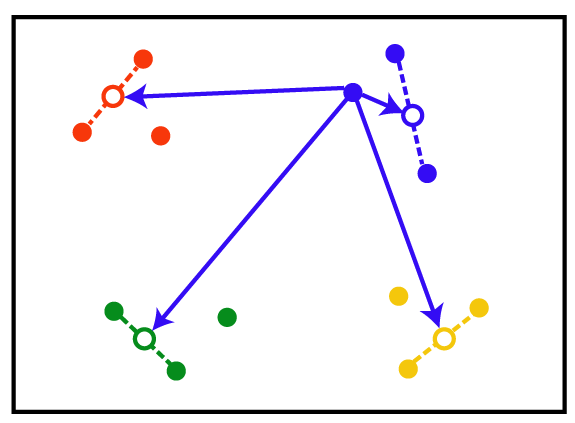}
        \caption{(Angular)Prototypical}
        \label{fig:3}
        \end{subfigure}
        \begin{subfigure}
        \centering
        \includegraphics[width=\textwidth]{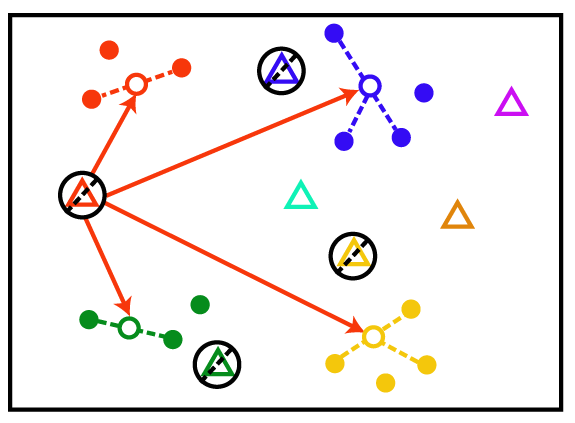}
        \caption{MP Regulator(Eq. \ref{l2})}
        \label{maskproxy_l2}
        \end{subfigure}%
    \end{minipage}%
    \hfil
    \begin{minipage}{.24\textwidth}
        \begin{subfigure}
        \centering
        \includegraphics[width=\textwidth]{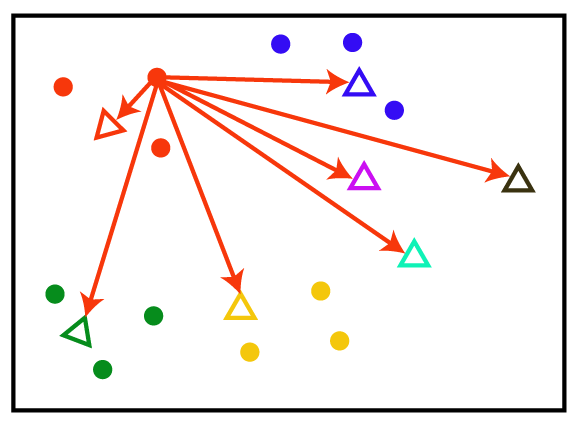}
        \caption{ProxyNCA}
        \label{proxynca}
        \end{subfigure}
        \begin{subfigure}
        \centering
        \includegraphics[width=\textwidth]{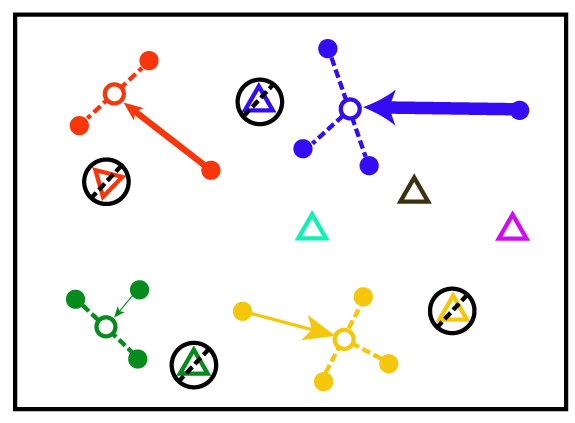}
        \caption{MMP Positive Pairs}
        \label{mmp_hardness}
        \end{subfigure}%
     
    \end{minipage}
\caption{\small Each color represents a unique class. Solid circle point is the entity while the triangle with a specific color is the proxy of the corresponding class. The hollow circle connected with solid circles by dash line represents the centroid computed by averaging those entities. For (Angular) Prototypical or MP (MMP), one embedding is reserved as a query and other embeddings are applied to compute the centroid regarding each class. No additional identities are incorporated for Contrastive, Triplet and (Angular) Prototypical. For proxy-based losses, all classes are included. The masked proxy is the triangle surrounded by a black circle and \textit{masked} by a dash line, shown in Fig. (\ref{mp_l1}) (\ref{maskproxy_l2}) (\ref{mmp_hardness}). For MP l1 loss, we compute the distance between the query and all centroids and the distance between the query and unmasked proxies, as shown in Fig. \ref{mp_l1}. For MP regulator, we compute the distance between the proxy and all centroids for all masked proxies, as shown in Fig. \ref{maskproxy_l2}. The illustration of MMP l1 loss is almost the same as that of MP l1 loss, except that each positive pair is assigned a weight which is correlated to the distance of this positive pair, which is shown by the width of lines in Fig. \ref{mmp_hardness}.}
\label{fig:images}
\end{figure*}

\subsection{Masked Proxy (MP)}
As seen in Equations \ref{nca} and \ref{anchor}, both Proxy NCA and Proxy Anchor only deal with entity-proxy pairs. In our proposed loss, both entity-proxy pairs and entity-entity pairs are taken into consideration. Specifically, for all classes represented within a minibatch we will compute entity-based distances, and compute entity-proxy distances for the remaining classes. In the following discussion, all computations are over a minibatch.

Let $L_i$ denote the class label of $x_i$ and $L_{p_i}$ denote the corresponding class label of proxy $p_i$. $L_M$ is the label set of classes represented in the minibatch. $c_{L_i}$ is the centroid \cite{Prototypical} of embeddings whose labels are $L_i$. $N_{L_i}$ is the number of instances in the minibatch with labels are $L_i$. Similar to \cite{GE2E,Prototypical, metric}, we reserve one instance from each class in the minibatch as a ``query'' for the class embedding as query for each class, and we define $X_Q$ as the query set for the minibatch, without duplicates. The centroids can be represented as follows:
\begin{equation}
    c_{L_i} = \frac{1}{N_{L_i}-1}  \sum_{\substack{j\\
                  j\neq i,
                  L_j=L_i}}x_j
\end{equation}
For each query instance in $X_Q$, we compose the following loss:
\begin{equation}\label{eq:maskproxy}
    l(x_i)=-\mathrm{\log}(\frac{e^{s(x_i,c_{L_i})}}{\sum\limits_{\substack{L_j\neq L_i \\ L_j\in L_M}}e^{ s(x_i,c_{L_j})}+\sum\limits_{\substack{p_k\\
        L_{p_k}\notin L_M
        }
}e^{s(x_i,p_k)}})
\end{equation}
where
\begin{equation} \label{distance}
    s(u,v) = \alpha ((u^Tv)-\beta)
\end{equation}
Both embeddings and proxies are normalized by length. $\alpha$ and $\beta$ are learnable smoothing factor and bias.

The numerator in Equation \ref{eq:maskproxy} considers the similarity of the instance to its own proxy. The first term in the denominator invokes the similarity of the instance to the {\em mini-batch centroids} for all classes in the minibatch, and is effectively entirely entity-based. The second term refers to the similarity of the instance to the proxies of all classes that are not in the minibatch and is proxy-based. The loss objective over the entire minibatch is formulated as follows:
\begin{equation}\label{eq:minibatchmaskproxy}
    l_1=\frac{1}{\left | X_Q \right |}\sum\limits_{x_i \in X_Q}l(x_i)
\end{equation}

Equation \ref{eq:minibatchmaskproxy} does not refer to the proxies of classes represented in the minibatch. In order to be able to also update these proxies classes, we include a {\em Mask Proxy Regulator} (MPR), which minimizes the distance between these proxies and the centroid for their class, while maximizing their separation to other centroids: 
\begin{equation}\label{l2}
    l_2=-\frac{1}{\left | L_M \right |}\sum\limits_{\substack{c_{L_i}\\
    L_i \in L_M, L_{p_i}=L_i}}\mathrm{\log}(\frac{e^{s(c_{L_i},p_i)}}{\sum\limits_{\substack{L_j\neq L_i}}e^{s(c_{L_j},p_i)}})
\end{equation}
Mask Proxy loss can be defined as the weighted summation of Equations \ref{eq:minibatchmaskproxy} and \ref{l2}:
\begin{equation}
    l_{mp}=l_1+\lambda_{mp} l_2
\end{equation}
where $\lambda_{mp}$ is balancing factor, which is set to 0.5 in this paper.

\subsection{Multinomial Masked Proxy (MMP)}
Multinomial-based loss functions can leverage both {\em multi-similarity} among pairs \cite{MSLOSS} and hardness of positive pairs \cite{Proxyanchor} as compared to softmax-based functions. The multinomial form of Equation \ref{eq:minibatchmaskproxy} can be formulated as follows:
\begin{equation}\label{l1M}
    \begin{split}
        l_{1m}&=\mathrm{\log}(1+\sum\limits_{x_i \in X_Q}e^{-s(x_i,c_{L_i})})\\&+\frac{1}{\left | X_Q \right |}\sum\limits_{x_i \in X_Q}\mathrm{\log}(1+\sum\limits_{\substack{L_j\neq L_i \\ L_j\in L_M}}e^{ s(x_i,c_{L_j})})\\&+\frac{1}{\left | X_Q \right |}\sum\limits_{x_i \in X_Q}\mathrm{\log}(1+\sum\limits_{\substack{p_k\\
        L_{p_k}\notin L_M
        }
    }e^{s(x_i,p_k)})
    \end{split}
\end{equation}
In Mask Proxy loss, every positive entity-centroid pair contributes equally to the loss objective, which can be illustrated through Eq. \ref{derivative}. 
\begin{equation} \label{derivative}
    \frac{\partial l_{MP}}{\partial s(x_i, c_{L_i})}=\frac{\partial l_1}{\partial s(x_i, c_{L_i})}=-\frac{1}{\left | X_Q \right |}
\end{equation}
However, multinomial loss will assign a dynamic weight to each positive pair. As shown in Equation  \ref{derivative_mmp}, this dynamic weight is dependent on the pair itself, meaning the harder positive pairs with lower similarity would contribute more to the loss objective. 
\begin{equation} \label{derivative_mmp}
    \frac{\partial l_{1m}}{\partial s(x_i, c_{L_i})}=-\frac{e^{-s(x_i,c_{L_i})}}{1+\sum\limits_{x_i \in X_Q}e^{-s(x_i,c_{L_i})}}
\end{equation}
The expression of MMP can be formulated as follows:
\begin{equation}
    l_{mmp}=l_{1m}+\lambda_{mmp} l_2
\end{equation}
Where the regulator $l_2$ is kept as its original form and $\lambda_{mmp}$ is a balancing factor. $\lambda_{mmp}$ is set to 0.5 in this paper.

A comparison of the different metric learning objectives described in this section and other standard metric learning objectives is shown in Fig. \ref{fig:images}.



\subsection{Training Complexity}
We denote $N, B, P$ as training size, batch size and proxy size respectively. For each data point in Eq. \ref{eq:minibatchmaskproxy}, we compute distances over C pairs, which gives rise to a training complexity of $O(NP)$. In Eq. \ref{l2}, complexity should be $O(P^2)$ to visit all possible proxy-centroid pairs.  Hence, the final training complexity should be within $O(NP+P^2) \sim O(NP)$, The same result can be derived for MMP. Table \ref{table1} is a comparison of training complexity over different loss functions.

\begin{table}[h!]
\centering
  \begin{threeparttable}
    \begin{tabular}{ |c|c| } 
     \hline
     Loss Criterion & Training Complexity\\ 
     \hline
     Triplet & $O(N^3)$\\ 
     Semi-hard Triplet \cite{Semihard} & $O(\frac{N^3}{B^2})$\\ 
     Smart Triplet \cite{smart_triplet} & $O(N^2)$\\  
     (Angular) Prototypical \cite{metric} & $O(N^2)$\tnote{*}\\
     Proxy NCA \cite{Proxy-NCA} & $O(NP)$\\  
     Proxy Anchor \cite{Proxyanchor} & $O(NP)$\\
     \textbf{MP/MMP} & $O(NP)$\\ 
     \hline
    \end{tabular}
     \caption{Training Complexity Comparison}
     \label{table1}
     
    \begin{tablenotes}
   \item[*] \small Here we discuss the case where number of data points is 2 for each class within the mini-batch.  
  \end{tablenotes}
  \end{threeparttable}
  \vspace{-0.6cm}
\end{table}

\section{Experiments}
\begin{table*}[!ht]
\centering
    \begin{tabular}{ |c|c|c|c|c|} 
     \hline
     Loss Function & Hyper-parameters& EER\%(E1, $\tau$=2s,B=400)& EER\%(E2,$\tau$=2s,B=800)&EER\%(E3,$\tau$=4s,B=400)\\ 
     \hline
     Triplet\cite{metric} & m=0.1 & 2.50$\pm$0.06&2.49$\pm$0.08&2.50$\pm$0.07\\ 
     ProtoTypical\cite{metric} & M=2 &2.37$\pm$0.10&2.34$\pm$0.08&2.32$\pm$0.02\\
     GE2E\cite{metric} & M=3 & 2.53$\pm$0.03&2.51$\pm$0.03&2.50$\pm$0.01\\ 
     Angular Prototypical\cite{metric} & M=2& 2.31$\pm$0.05&2.21$\pm$0.03&2.26$\pm$0.05\\ 
     Proxy NCA & $\setminus$ &2.42$\pm$0.07&2.42$\pm$0.05&2.40$\pm$0.04\\
     Proxy Anchor & m=0.15,s=50 & 2.39$\pm$0.03&2.40$\pm$0.02&2.39$\pm$0.01\\ 
     MP(ours) & $\lambda$=0.3& \textbf{2.08$\pm$0.03}&\textbf{2.04$\pm$0.05}&\textbf{2.05$\pm$0.03}\\ 
     MMP(ours) & $\lambda$=0.3 &\textbf{2.06$\pm$0.03}&\textbf{2.02$\pm$0.04}&\textbf{2.02$\pm$0.02}\\ 
     MP-Balance(ours) & M=2,$\lambda$=0.3 & \textbf{2.03$\pm$0.01}&\textbf{1.97$\pm$0.03}&\textbf{1.99$\pm$0.01}\\ 
     MMP-Balance(ours) & M=2,$\lambda$=0.3 & \textbf{1.99$\pm$0.02}&\textbf{1.95$\pm$0.03}&\textbf{1.93$\pm$0.01}\\ 
     \hline
    \end{tabular}
     \caption{Evaluation on VoxCeleb1 test set}
     \label{table2}
     \vspace{-0.7cm}
\end{table*}
\vspace{-0.3cm}
\subsection{Dataset}
We use the VoxCeleb dataset \cite{voxCeleb,voxceleb2}. We train on VoxCeleb2 dev set which contains 5994 identities and test on VoxCeleb1 test set which contains 40 identities.
\vspace{-0.3cm}
\subsection{Model and Evaluation Scoring Method}
We apply the backbone (Thin-ResNet34+SAP) proposed in \cite{metric} as baseline model, where SAP is self-attentive pooling \cite{SAP}. We adopt Equal Error Rate(EER) as evaluation metric, where the similarity score is computed using the same method with \cite{voxceleb2, metric}: 10 features are sampled using a specific window size for each utterance in a pair, then the mean of $10\times 10$ Euclidean distances are computed as the distance measurement. Note that these 10 features cover the full utterance. That is to say, during testing, the full utterance rather than a fixed-length sample from the raw audio is evaluated. Details for evaluation method can be found in \cite{voxceleb2, metric}. 
\vspace{-0.3cm}
\subsection{Implementation Details}
\label{implementation}
The experiments are performed on Nvidia Tesla V100 platform using PyTorch. We denote $\tau$ as audio duration. We apply fixed $\tau=2s$ and $\tau=4s$ random segments respectively during training while using the full utterance during testing \cite{metric, voxceleb2}. 
We employ the Mel-Spectrogram feature with 40 frequency channels. In accordance with \cite{metric}, we adopt the window size of 25ms and step size of 10ms. Sampling rate is set as 16k. 

We experiment with Proxy NCA, Proxy Anchor, MP and MMP on Thin-ResNet34. When experimenting with Proxy Anchor, different sets of hyper-parameters are explored. To be specific, margin is varied from 0.1 to 0.5 with an increment of 0.1 and smoothing factor is varied from 10 to 70 with an increment of 10. The balancing factor $\lambda$ is varied from 0.1 to 1 with an increment of 0.1. While training with MP and MMP, the initial margin $\beta$ is set to 0.1 and smoothing factor $\alpha$ is set to 10.  Inspired by the idea that balanced training samples matter \cite{GE2E,Prototypical, metric}, we also adopt this manner in MP and MMP. Concretely, the number of samples for each class is a fixed value $M$ which is a hyper-parameter. In this experiment, we set $M$ as 2 based on the fact that \emph{fewer shot} learning matters in Angular Prototypical loss \cite{metric}. We call this manner \textbf{MP-Balance} for Mask Proxy and \textbf{MMP-Balance} for Multinomial Mask Proxy. We adopt the *expected batch size* $B$ of 800 and 400\footnote{For MP/MMP-Balance, input size is 2 $\times$ {[$batchsize$, feature size]} and *expected batch size* is $batchsize \times 2$, which is consistent with \cite{metric}. For MP and MMP, input size is [$\sum_{i=1}^{batchsize}M_i$, feature size], where $M_i$ is number of features for a certain class $i$. $M_i$ is randomly determined to be 2 or 3, and  $\sum_{i=1}^{batchsize}M_i=batchsize\times2.5$ is the *expected batch size*. E.g. for *expected batch size* 800, the $batchsize$ is set to 320.} respectively. We apply SGD as optimizer with a starting learning rate of 0.2 and ReduceLROnPlateau as learning rate scheduler with a factor of 0.8, patience of 3 and $EER\%$ as metric. 

Based on the aforementioned statement, three experiments are performed and we denote them as $\textbf{E1}(\tau=2s, B=400)$, $\textbf{E2}(\tau=2s, B=800)$, $\textbf{E3}(\tau=4s, B=400)$ respectively. We also duplicate the experiments in \cite{metric} using Triplet, Prototypical, GE2E and Angular Prototypical respectively.  To analyze the training speed over different loss objectives, in each epoch, we compute the $EER\%$ on test set and record the results in Fig. \ref{convergence}. Based on our experiments, there is no significant difference on training convergence over $E1$, $E2$ and $E3$. Thus we just take $E1$ into consideration in Fig. \ref{convergence}. The final results are displayed in Table \ref{table2}. The hyper-parameters in both Table. \ref{table2} and Fig. \ref{convergence} are the best ones that we have explored.

\subsection{Discussion}
 \textbf{Results Analysis}. Table \ref{table2} presents the EER achieved using different loss functions under certain settings. The performance for Proxy NCA and Proxy Anchor are similar to that of Prototypical. 
 Based on the current best smoothing factor and margin, Proxy Anchor reaches slightly lower EER than Proxy NCA. The potential reason is that Proxy Anchor indirectly leverages the fine-grained data relationship. 

The proposed loss functions including Mask Proxy and Multinomial Mask Proxy outperform the existing state-of-the-art Angular Prototypical loss objective.
Based on these results, we speculate that both fine-grained data-to-data connection and number of classes incorporated in the mini-batch are the major contributing factors for a lower EER achieved by the proposed loss functions. 
\vspace{-0.6cm}
\begin{figure}[!ht]
\centering
\includegraphics[scale=0.24]{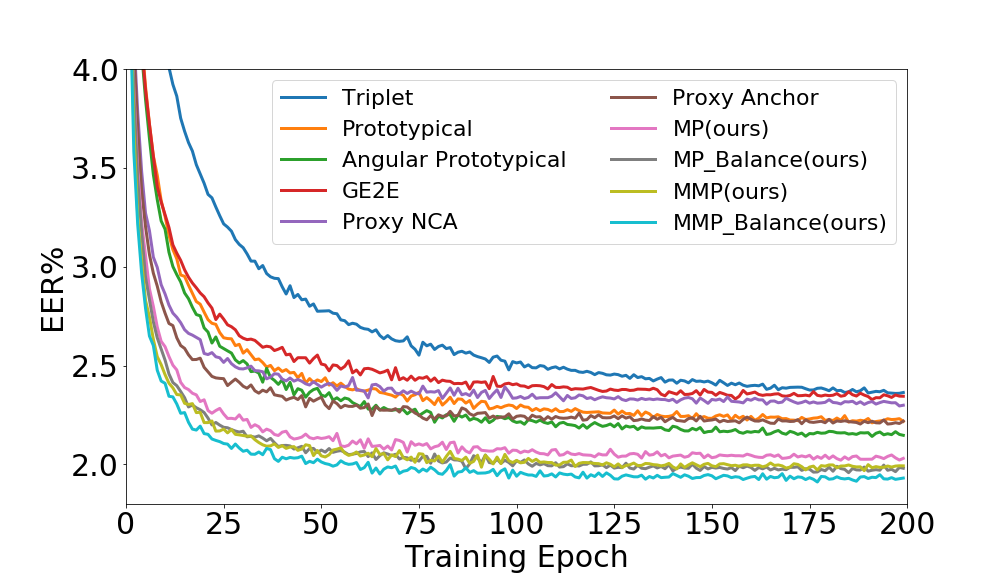}
\caption{\small EER on the test set - VoxCeleb1 dataset using different loss objectives over epochs. Details can be found in Sec.  \ref{implementation}}
  \label{convergence}
  \vspace{-0.3cm}
\end{figure}

\textbf{Ablation Study}
Based on the results in Table. \ref{table2}, balanced data inputs always result in lower EER for both MP and MMP. Both larger batch size $B$ and larger training audio duration $\tau$ give lower EER for MP, MMP, MP-Balance and MMP-Balance. These conclusions hold for all three experiments $E1$, $E2$ and $E3$. 

\textbf{Training Complexity and Speed}.
Based on our experiments, both training audio duration and batch size make no difference in the training speed. To be concise, we just choose the training process of $E1$ in Fig. \ref{convergence} for the illustration of training speed over different loss functions. It is observed that Triplet converges the most slowly due to its largest training complexity as shown in Table. \ref{table1}. Other non-proxy-based losses converge faster than Triplet due to its smaller training complexity. Proxy-based losses converge fastest. Note that though Proxy NCA and Proxy Anchor give a larger final EER in comparison to that of Angular Prototypical, they still converge faster than Angular Prototypical. Our proposed loss functions give the best EER as well as training speed.


\section{Conclusion}
The proposed loss objectives in this paper achieve state-of-the-art speaker verification performance on the VoxCeleb dataset. To the best of our knowledge, it is the first work that applies proxy-based losses on the task of speaker verification. The proposed Masked Proxy losses leverage real data-to-data relationships more than other proxy-based objectives in a novel manner. Our future work will focus on self-supervised and low-resource speaker recognition.

\bibliographystyle{IEEEtran}

\bibliography{main}


\end{document}